\documentclass[onecolumn,journal,noadjust]{IEEEtran}
\usepackage{amsmath,amsfonts}
\usepackage{algorithmic}
\usepackage{algorithm}
\usepackage{array}
\usepackage[caption=false,font=normalsize,labelfont=sf,textfont=sf]{subfig}
\usepackage{textcomp}
\usepackage{stfloats}
\usepackage{url}
\usepackage{verbatim}
\usepackage{graphicx}
\usepackage{epstopdf} 
\usepackage{cite}

\usepackage{amssymb}
\usepackage{dsfont}
\usepackage{float}
\usepackage{graphicx}
\newtheorem{lemma}{Lemma}

\newtheorem{definition}{Definition}

\newtheorem{theorem}{Theorem}
\newtheorem{remark}{Remark}

\usepackage{xcolor}
\usepackage{mathrsfs}
\usepackage{caption}
\usepackage{multirow}
\usepackage{makecell}

\usepackage [autostyle, english = american]{csquotes}
\MakeOuterQuote{"}

\usepackage{amsmath}

\begin{document}

\title{Improved Channel Coding Performance Through Cost Variability}

\author{Adeel Mahmood and Aaron B.~Wagner\\
School of Electrical and Computer Engineering, Cornell University}



\maketitle

\begin{abstract}
Channel coding for discrete memoryless channels (DMCs) with mean and variance cost constraints has been recently introduced. We show that there is an improvement in coding performance due to cost variability, both with and without feedback. We demonstrate this improvement over the traditional almost-sure (per-codeword) cost constraint that prohibits any cost variation above a fixed threshold. Our result simultaneously shows that feedback does not improve the second-order coding rate of simple-dispersion DMCs under the almost-sure cost constraint. This finding parallels similar results for unconstrained simple-dispersion DMCs, additive white Gaussian noise (AWGN) channels and parallel Gaussian channels.      
\end{abstract}


\section{Introduction}

Channel coding is a fundamental problem focused on the reliable transmission of information over a noisy channel. Information transmission with arbitrarily small error probability is possible at all rates below the capacity $C$ of the channel, if the number $n$ of channel uses (also called the blocklength) is permitted to grow without bound~\cite{korner1}. Likewise, for any fixed error probability $\epsilon > 0$, the maximum transmission rate can get arbitrarily close to the capacity as $n \to \infty$. The second-order coding rate (SOCR) (\cite{strassen,5290292,5452208, 9099482,7447062,6816070}) quantifies the $O(n^{-1/2})$ convergence to the capacity.  

In many practical scenarios, the channel input is subject to some cost constraints which limit the amount of resources that can be used for transmission. With a cost constraint present, the role of capacity is replaced by the capacity-cost function \cite[Theorem 6.11]{korner1}. One common form of the cost constraint is the almost-sure (a.s.) cost constraint (\cite{5290292,6767457}) which bounds the time-average cost of the channel input $X^n$ over all messages, realizations
of any side randomness, channel noise (if there is feedback), etc.:  
\begin{align}
    \label{eq:constraint:as}
    \frac{1}{n} \sum_{i = 1}^n c(X_i) \le \Gamma \quad \text{almost surely,}
\end{align}
where $c(\cdot)$ is the cost function. Under the almost-sure (a.s.) cost constraint, the optimal first-order coding rate is the capacity-cost function, the strong converse holds \cite[Theorem 6.11]{korner1}, and the optimal SOCR is also known \cite[Theorem 3]{5290292}. 

The a.s.\ cost constraint is quite unforgiving, never allowing the
cost to exceed the threshold under any circumstances.
Our first result (Theorem \ref{thm2}) shows that the SOCR can be strictly improved by merely allowing the cost to fluctuate above the
threshold in a manner consistent with a noise process, i.e.,
the fluctuations have a variance of $O(1/n)$.
Our second result (Theorem \ref{thm}) shows that the a.s.\ cost framework does not allow feedback improvement to SOCR for simple-dispersion\footnote{See Definition \ref{defsimpl_disp}.} DMCs. This again contrasts with the scenario where random fluctuations with variance $O(1/n)$ above the threshold
are allowed, as shown in \cite[Theorem 3]{mahmood2024channel}. This highlights the important role cost variability plays in enabling feedback mechanisms to improve coding performance.

These findings raise the question of whether it is necessary to
impose a constraint as stringent as~(\ref{eq:constraint:as}).
Cost constraints in communication systems are typically imposed to achieve goals such as operating circuitry in the linear regime, minimizing power consumption, and reducing interference with other terminals. It is worth noting that these goals do not always necessitate the use of the strict a.s.\ cost constraint. For example, the expected cost constraint is often used in wireless communication literature (see, e.g., \cite{414651, 641562, 737514, 5374062}) because it allows for a dynamic allocation of power based on the current channel state. The expected cost constraint bounds the cost averaged over time and the ensemble:
\begin{align}
    \label{eq:constraint:mean}
    \mathbb{E}\left [ \frac{1}{n} \sum_{i = 1}^n c(X_i) \right ]  \le \Gamma.
\end{align}

Yet, if the a.s.\ constraint is too strict, the expectation constraint
is arguably too weak.
The expectation constraint allows highly non-ergodic use of power, 
as shown in Section~\ref{more_details},
which is problematic both from the vantage points of operating circuitry
in the linear regime and interference management.

The $O(1/n)$ variance allowance is a feature of a new cost formulation, referred to as \emph{mean and variance} cost constraints in \cite{mahmood2024channel}. This formulation replaces~(\ref{eq:constraint:as}) with the following conditions: 
\begin{align}
	\mathbb{E}\left [ \frac{1}{n} \sum_{i = 1}^n c(X_i) \right ] &  \le \Gamma, \label{exp01}\\
	\text{Var}\left ( \frac{1}{n} \sum_{i = 1}^n c(X_i) \right )  & \le \frac{V}{n}.
	\label{var01}
\end{align}
The mean and variance cost constraints were introduced as a relaxed version of the a.s.\ cost constraint that permits a small amount of stochastic
fluctuation above the threshold $\Gamma$ while providing an
ergodicity guarantee.
Consider a random channel codebook whose codewords satisfy $(\ref{exp01})$ with equality. For a given input $x^n$, define an ergodicity metric $\mathcal{E}_m$ as 
\begin{align}
	\mathcal{E}_m(x^n) := \frac{\max\left( \frac{1}{n} \sum_{i = 1}^n c(x_i) - \Gamma, 0 \right) }{\Gamma}. \label{h23} 
\end{align}
The definition in $(\ref{h23})$ only penalizes cost variation above the threshold and normalizes by the mean cost $\Gamma$. Let $\alpha > 0$ be the desired ergodicity parameter. We say that a transmission $x^n$ is \emph{$\alpha$-ergodic} if $\mathcal{E}_m(x^n) \leq \alpha$. Let $\beta$ be the desired uncertainty parameter. We say that a random codebook is \emph{$(\alpha, \beta)$-ergodic} if $\mathbb{P}\left(\mathcal{E}_m(X^n) \leq \alpha \right) \geq 1 - \beta$, where $X^n$ is a random transmission from the codebook.     

Under the mean and variance cost formulation, we have $\mathbb{P}\left(\mathcal{E}_m(X^n) \leq \alpha \right) \geq 1 - \beta$ if
\begin{align}
    n \geq n_c := \frac{V}{\beta \alpha^2 \Gamma^2}, \label{criticaln}
\end{align}
where we call $n_c$ the \emph{critical blocklength}. Thus, the critical blocklength specifies a blocklength of a channel code for and above which a transmission behaves ergodically with high probability. 
For fixed $\alpha$, $\beta$ and $\Gamma$, the parameters $n_c$ and $V$ have a one-to-one relation given in $(\ref{criticaln})$.
Note that with an
expectation-only constraint, we effectively have $V = \infty$,
so the transmission is not guaranteed to be ergodic at any blocklength.
Furthermore, unlike the expected cost constraint, the mean and variance cost formulation: 
\begin{itemize}
    \item allows for a strong converse \cite[Theorem 77]{Polyanskiy2010}, \cite{mahmood2024channel},
    \item allows for a finite second-order coding rate \cite{mahmood2024channel}, 
    \item does not allow blasting power on errors in the feedback case \cite{1523364}. 
\end{itemize}

The results of this paper also have significance in the context of previous works. Our result in Theorem \ref{thm} extends the previously known result that feedback does not improve the second-order performance for simple-dispersion DMCs without cost constraints \cite{9099482}. It is also similar to the result in \cite{7282467} that feedback does not improve the second-order performance for AWGN channels.

To summarize, let 
\begin{itemize}
    \item $r_{\text{a.s.}}(\epsilon, \Gamma)$ denote the optimal SOCR with the a.s. cost constraint without feedback,
    \item $r_{\text{a.s.,fb}}(\epsilon, \Gamma)$ denote the optimal SOCR with the a.s. cost constraint with feedback,
    \item $r_{\text{m.v.}}(\epsilon, \Gamma, V)$ denote the optimal SOCR with the mean and variance cost constraints without feedback and  
    \item $r_{\text{m.v.,fb}}(\epsilon, \Gamma, V)$ denote the optimal SOCR with the mean and variance cost constraints with feedback,  
\end{itemize}
for DMC codes operating with average error probability of at most $\epsilon \in (0, 1)$. The mathematical definitions of the four quantities above are given in Section \ref{main_results}. Under broad assumptions covering many practical channels, we establish the following hierarchy of second-order coding rates:
\begin{align*}
    r_{\text{m.v.,fb}}(\epsilon, \Gamma, V) > r_{\text{m.v.}}(\epsilon, \Gamma, V) > r_{\text{a.s.,fb}}(\epsilon, \Gamma) = r_{\text{a.s.}}(\epsilon, \Gamma).     
\end{align*}
The assumptions underlying the above hierarchy are summarized toward the end of Section \ref{main_results}.

\subsection{Related Work}

The second- and third-order asymptotics for DMCs and AWGN channels with the a.s. cost constraint in the non-feedback setting have been characterized in \cite{5290292} (second-order), \cite{7055296} (third-order, DMCs) and \cite{7056434} (third-order, AWGN channels). The second-order asymptotics in the feedback setting of DMCs that are not simple-dispersion are studied in \cite{9099482} without cost constraints. There are more feedback results available for AWGN channels compared to DMCs under the a.s. cost constraint. For example, the result in \cite{7282467} also addresses the third-order performance with feedback while \cite{8012458} gives the result that feedback does not improve the second-order performance for parallel Gaussian channels. The second-order performance for the AWGN channel with an expected cost constraint is characterized in \cite{7156144}. Table~\ref{prevworks} summarizes these results across different settings in channel coding.

\begin{table*}[h]
\begin{minipage}{\textwidth} 
\begin{center}
\setcellgapes{2pt}
\makegapedcells
\begin{tabular}{|c|c|c|c|c|c|}
\hline
    Paper & Channel \ & Performance  &
        Cost Constraint\ & Feedback & Non-feedback \\
        \hline
    Hayashi \cite{5290292} & DMC, AWGN & 2nd order & a.s. & No  & Yes
        \\
        Kostina and Verdú \cite{7055296}  & DMC & 3rd order & a.s.  & No & Yes \\
        Tan and Tomamichel \cite{7056434} & AWGN & 3rd order & a.s.  & No & Yes \\
        Wagner, Shende and Altuğ \cite{9099482}  & DMC & 2nd order & none & Yes & No \\
        Fong and Tan \cite{7282467} & AWGN & 2nd and 3rd order & a.s.  & Yes & No \\
    Mahmood and Wagner \cite{mahmood2024channel} & DMC & 2nd order  & mean and variance & Yes & Yes \\
    Mahmood and Wagner \cite{mahmood2025channelcodinggaussianchannels} & AWGN & 2nd order  & mean and variance & No & Yes \\
    This paper  & DMC & 2nd order & mean and variance, a.s. & Yes & Yes \\
    Polyanskiy \cite[Th. 78]{Polyanskiy2010} & Parallel AWGN  & 2nd order & a.s. & No & Yes \\
    Fong and Tan \cite{8012458} & Parallel AWGN  & 2nd order & a.s. & Yes & No \\
    Polyanskiy \cite[Th. 77]{Polyanskiy2010} & AWGN & 1st order & expected cost & No & Yes\\
    Yang et al. \cite{7156144} & AWGN & 2nd order & expected cost & No & Yes\\
    Polyanskiy at al. \cite[Th. 54]{5452208} &  AWGN & 2nd order & expected cost\footnote{This result is stated for the maximal probability of error framework while Yang et al. \cite[Theorem 1]{7156144} applies for the average error probability.} & No  & Yes
        \\
    \hline
\end{tabular}
\end{center}
\caption{Relevant results across different settings in channel coding.}
\label{prevworks}
\end{minipage}
\end{table*}

Our proof technique for Theorem \ref{thm} is more closely aligned with that used in \cite{9099482} for DMCs than in \cite{7282467} for AWGN channels. Both proofs show converse bounds with feedback that match the previously known non-feedback achievability results for DMCs and AWGN channels, respectively. A common technique used in both converse proofs is a result from binary hypothesis testing, which is used in the derivation of Lemma \ref{qandp2} in our paper and a similar result in \cite[(17)]{7282467}. We then proceed with the proof by using a Berry-Esseen-type result for
bounded martingale difference sequences whereas \cite{7282467} uses the usual Berry-Esseen theorem by first showing equality in distribution of the information density with a sum of i.i.d. random variables.

\section{Preliminaries}

Let $\mathcal{A}$ and $\mathcal{B}$ be finite input and output alphabets, respectively, of the DMC $W$, where $W$ is a stochastic matrix from $\mathcal{A}$ to $\mathcal{B}$. For a given sequence $x^n \in \mathcal{A}^n$, the $n$-type $t = t(x^n)$ of $x^n$ is defined as
\begin{align*}
    t(a) &= \frac{1}{n} \sum_{i=1}^n \mathds{1}(x_i = a)
\end{align*}
for all $a \in \mathcal{A}$, where $\mathds{1}(.)$ is the standard indicator function. For a given sequence $x^n \in \mathcal{A}^n$, we will use $t(x^n)$ or $P_{x^n}$ to denote its type. Let $\mathcal{P}_n(\mathcal{A})$ be the set of $n$-types on $\mathcal{A}$. For a given $t \in \mathcal{P}_n(\mathcal{A})$, $T^n_{\mathcal{A}}(t)$ denotes the type class, i.e., the set of sequences $x^n \in \mathcal{A}^n$ with empirical distribution equal to $t$. For a random variable $Z$, $||Z||_\infty$ denotes its
essential supremum (that is, the infimum of those numbers $z$ such that $\mathbb{P}(Z \leq z) = 1$). We will write $\log$ to denote logarithm to the base $e$ and $\exp(x)$ to denote $e$ to the power of $x$. The cost function is denoted by $c(\cdot)$ where $c: \mathcal{A} \to [0, c_{\max}]$ and $c_{\max} > 0$ is a constant. Let $\Gamma_0 = \min_{a \in \mathcal{A}} c(a)$. Let $\Gamma^*$ denote the smallest $\Gamma$ such that the capacity-cost function  $C(\Gamma)$ is equal to the unconstrained capacity. We assume $\Gamma^* > \Gamma_0$ and $\Gamma \in (\Gamma_0, \Gamma^*)$ throughout the paper. For $\Gamma \in (\Gamma_0, \Gamma^*)$, the capacity-cost function is defined as 
\begin{align}
    C(\Gamma ) &= \max_{\substack{P \in \mathcal{P}(\mathcal{A})\\
    c(P) \leq \Gamma} } I(P, W), \label{main_form}
\end{align}
where $c(P) := \sum_{a \in \mathcal{A}} P(a) c(a)$. The function $C(\Gamma)$ is strictly increasing and differentiable \cite[Problem 8.4]{korner1} in the interval $(\Gamma_0, \Gamma^*)$. For a given $x^n \in \mathcal{A}^n$, we define 
\begin{align*}
    c(x^n) := \frac{1}{n} \sum_{i=1}^n c(x_i).
\end{align*}

Let $\Pi_{W, \Gamma}^*$ be the set of all capacity-cost-achieving distributions, i.e., the set of maximizing distributions in $(\ref{main_form})$. For any $P^* \in \Pi_{W, \Gamma}^*$, let $Q^* = P^*W$ be the marginal distribution on $\mathcal{B}$. Note that the output distribution $Q^*$ is always unique, and without loss of generality, $Q^*$ can be assumed to satisfy $Q^*(b) > 0$ for all $b \in \mathcal{B}$ \cite[Corollaries 1 and 2 to Theorem 4.5.1]{gallager1968}. 

The following definitions will remain in effect throughout the paper: 
\begin{align*}
\nu_{a} &:= \text{Var}\left( \log \frac{W(Y|a)}{Q^*(Y)} \right),\quad  \text{ where } Y \sim W(\cdot|a),\\
\nu_{\max} &:= \max_{a \in \mathcal{A}} \nu_a,\\
    i(a, b) &:= \log \frac{W(b|a)}{Q^*(b)}.
\end{align*}
\begin{definition}[cf. \cite{9099482}]
    A DMC $W$ is called simple-dispersion at the cost $\Gamma \in (\Gamma_0, \Gamma^*)$ if 
    \begin{align*}
        \min_{P^* \in \Pi_{W, \Gamma}^*} \sum_{a \in \mathcal{A}} P^*(a) \nu_{a} = \max_{P^* \in \Pi_{W, \Gamma}^*} \sum_{a \in \mathcal{A}} P^*(a) \nu_{a}.
    \end{align*}
    \label{defsimpl_disp}
\end{definition}
We will only focus on simple-dispersion channels for a fixed cost $\Gamma \in (\Gamma_0, \Gamma^*)$ and thus define 
$$V(\Gamma) := \sum_{a \in \mathcal{A}} P^*(a) \nu_{a}$$
for any $P^* \in \Pi_{W, \Gamma}^*$.

With a blocklength $n$ and a fixed rate $R > 0$, let $\mathcal{M}_R = \{1, \ldots, \lceil \exp(nR) \rceil \}$ denote the message set. Let $M \in \mathcal{M}_R$ denote the random message drawn uniformly from the message set.   

\begin{definition}
An $(n, R)$ code for a DMC consists of an encoder $f$ which, for each message $m \in \mathcal{M}_R$, chooses an input $X^n = f(m) \in \mathcal{A}^n$, and a decoder $g$ which maps the output $Y^n$ to $\hat{m} \in \mathcal{M}_R$. The code $(f,g)$ is random if $f$ or $g$ is random. 
\label{defwocostwofeedback}
\end{definition}

\begin{definition}
An $(n, R)$ code with ideal feedback for a DMC consists of an encoder $f$ which, at each time instant $k$ ($1 \leq k \leq n$) and for each message $m \in \mathcal{M}_R$, chooses an input $x_k = f(m, x^{k-1}, y^{k-1}) \in \mathcal{A}$, and a decoder $g$ which maps the output $y^n$ to $\hat{m} \in \mathcal{M}_R$. The code $(f,g)$ is random if $f$ or $g$ is random.   
\label{defwocostwfeedback}
\end{definition}

\begin{definition}
An $(n, R, \Gamma)$ code for a DMC is an $(n, R)$ code such that $ c(X^n) \leq \Gamma$ almost surely, where the message $M \sim \text{Unif}(\mathcal{M}_R)$ has a uniform distribution over the message set $\mathcal{M}_R$. 
\label{defwcostwofeedback}
\end{definition}

\begin{definition}
    An $(n, R, \Gamma)$ code with ideal feedback for a DMC is an $(n,R)$ code with ideal feedback such that $c(X^n) \leq \Gamma$ almost surely, where the message $M \sim \text{Unif}(\mathcal{M}_R)$ has a uniform distribution over the message set $\mathcal{M}_R$. 
    \label{defwcostwfeedback}
\end{definition}

\begin{definition}
An $(n, R, \Gamma, V)$ code for a DMC is an $(n, R)$ code such that $\mathbb{E}\left [ \sum_{i=1}^n c(X_i) \right] \leq n \Gamma$ and $\text{Var}\left(\sum_{i=1}^n c(X_i) \right) \leq n V$, where the message $M \sim \text{Unif}(\mathcal{M}_R)$ has a uniform distribution over the message set $\mathcal{M}_R$. 
\label{defwcostwofeedback2}
\end{definition}

\begin{definition}
    An $(n, R, \Gamma, V)$ code with ideal feedback for a DMC is an $(n,R)$ code with ideal feedback such that $\mathbb{E}\left[\sum_{i=1}^n c(X_i)\right] \leq n\Gamma$ and $\text{Var}\left(\sum_{i=1}^n c(X_i) \right) \leq n V$, where the message $M \sim \text{Unif}(\mathcal{M}_R)$ has a uniform distribution over the message set $\mathcal{M}_R$. 
    \label{defwcostwfeedback2}
\end{definition}


Given $\epsilon \in (0, 1)$, define 
\begin{align*}
    M^*_{\text{fb}}(n, \epsilon,  \Gamma) := \max \{ \lceil \exp(nR) \rceil : \bar{P}_{\text{e,fb}}(n,R, \Gamma) \leq \epsilon   \},
\end{align*}
where $\bar{P}_{\text{e,fb}}(n,R,\Gamma)$ denotes the minimum average error probability attainable by any random $(n,R, \Gamma)$ code with feedback. Similarly, define \begin{align*}
    M^*(n, \epsilon,  \Gamma) := \max \{ \lceil \exp(nR) \rceil : \bar{P}_{\text{e}}(n,R, \Gamma) \leq \epsilon   \},
\end{align*}  
where $\bar{P}_{\text{e}}(n,R,\Gamma)$ denotes the minimum average error probability attainable by any random $(n,R, \Gamma)$ code without feedback. Define $M^*_{\text{fb}}(n, \epsilon, \Gamma, V)$ and $M^*(n, \epsilon, \Gamma, V)$ similarly for codes with mean and variance cost constraints.

\subsection{Expectation-only cost constraint \label{more_details}}

Under this cost formulation, the average cost of the codewords is constrained in expectation only:
\begin{align}
    \mathbb{E}\left [ \frac{1}{n} \sum_{i=1}^n c(X_i) \right ] &\leq \Gamma.  \label{exp_cost}
\end{align}
We now illustrate a codebook construction (adapted from \cite{7055296}) with an average error probability at most $\epsilon \in (0, 1)$ that meets the cost threshold $\Gamma$ according to $(\ref{exp_cost})$ but the cost of its codewords is non-ergodic, i.e., $\frac{1}{n} \sum_{i=1}^n c(X_i)$ does not converge to $\Gamma$. Consider a codebook $\mathcal{C}_n$ with rate $C(\Gamma) < R < C\left(\frac{\Gamma}{1- \epsilon}\right)$ whose average error probability $\epsilon_n \to 0$ and each of whose codewords has average cost equal to $\frac{\Gamma}{1- \epsilon}$. Such a codebook exists because $R < C\left(\frac{\Gamma}{1- \epsilon}\right)$. Assuming $\Gamma_0 = \min_{a \in \mathcal{A}} c(a) = 0$ without loss of generality, one could modify the codebook $\mathcal{C}_n$ by replacing an $\epsilon$-fraction of its codewords with the all-zero codeword. The modified codebook $\mathcal{C}_n'$ has average error probability at most $\epsilon_n' \to \epsilon$ and meets the cost threshold $\Gamma$ according to $(\ref{exp_cost})$. But $\frac{1}{n} \sum_{i=1}^n c(X_i)$ is either $0$ or $\frac{\Gamma}{1- \epsilon}$. This construction also shows that the strong converse does not hold under the expected cost constraint. 

The mean and variance cost constraints ensure that the average cost of the codewords concentrate around the cost threshold $\Gamma$, thereby disallowing codebook constructions with irregular or non-ergodic power consumption.

\section{Main Results \label{main_results}}

We prove coding performance improvement in terms of the second-order coding rate, although equivalent results in terms of the average error probability improvement can also be shown as in \cite[Theorems 1-3]{mahmood2024channel}. In the non-feedback case, $C(\Gamma)$ is the optimal first-order rate for DMCs with the a.s. cost constraint \cite[Theorem 6.11]{korner1} as well as the mean and variance cost formulation \cite[Theorems 1 and 2]{mahmood2024channel}, i.e., 
\begin{align}
    \lim_{n \to \infty} \frac{1}{n} \log M^*(n, \epsilon, \Gamma) &= C(\Gamma) \label{str1}\\
    \lim_{n \to \infty} \frac{1}{n} \log M^*(n, \epsilon, \Gamma, V) &= C(\Gamma) \label{str2}
\end{align}
for all $\epsilon \in (0, 1)$. The results $(\ref{str1})$ and $(\ref{str2})$ imply that the strong converse holds. We thus define the second-order rates with respect to the capacity-cost function $C(\Gamma)$ as follows: 
\begin{align*}
    r_{\text{a.s.}}(\epsilon, \Gamma) &:= \liminf_{n \to \infty} \frac{\log M^*(n, \epsilon, \Gamma) - n C(\Gamma)}{\sqrt{n}}\\
    r_{\text{m.v.}}(\epsilon, \Gamma, V) &:= \liminf_{n \to \infty} \frac{\log M^*(n, \epsilon, \Gamma, V) - n C(\Gamma)}{\sqrt{n}}.
\end{align*}
For the feedback case, we simply take the convention to define the SOCR with respect to $C(\Gamma)$ as follows:     
\begin{align*}
    r_{\text{a.s.,fb}}(\epsilon, \Gamma) &:= \liminf_{n \to \infty} \frac{\log M^*_{\text{fb}}(n, \epsilon, \Gamma) - n C(\Gamma)}{\sqrt{n}}\\
     r_{\text{m.v.,fb}}(\epsilon, \Gamma, V) &:= \liminf_{n \to \infty} \frac{\log M^*_{\text{fb}}(n, \epsilon, \Gamma, V) - n C(\Gamma)}{\sqrt{n}}.
\end{align*}
For the a.s. cost constraint, this convention is justified because from the result in Theorem \ref{thm}, $C(\Gamma)$ is the optimal first-order rate, in the analogous sense to $(\ref{str1})$, for DMCs with feedback. For DMCs without cost constraints, Shannon \cite[Theorem 6]{1056798} showed that feedback does not increase the capacity.

The SOCR is helpful to characterize the second-order term in the asymptotic expansion of the maximum achievable rate. Rigorously, second-order achievability results take the form  
$$\liminf_{n \to \infty} \frac{\log M^* - n C}{\sqrt{n}} \geq \alpha,$$
while second-order converse results take the form 
$$\limsup_{n \to \infty} \frac{\log M^* - n C}{\sqrt{n}} \leq \beta.$$
If $\alpha = \beta = r$, then one has the asymptotic expansion   
\begin{align*}
    \log M^* &= n C  + \sqrt{n} \,r + o(\sqrt{n}). 
\end{align*}

\subsection{Performance improvement for non-feedback codes}

From \cite[Theorem 3]{5290292}, we have $r_{\text{a.s.}}(\epsilon, \Gamma) = \sqrt{V(\Gamma)} \Phi^{-1}(\epsilon)$ for a simple-dispersion\footnote{The result in \cite[Theorem 3]{5290292} is not restricted to simple-dispersion DMCs.} DMC $W$. On the other hand, 
\cite[Theorems 1 and 2]{mahmood2024channel} proved that 
\begin{align}
    &r_{\text{m.v.}}(\epsilon, \Gamma, V) = \mbox{} \notag \\
    & \quad \quad \max \left \{r \in \mathbb{R} : \mathcal{K}\left(\frac{r}{\sqrt{V(\Gamma) }}, \frac{C'(\Gamma)^2 V}{V(\Gamma)} \right) \leq \epsilon \right \} \label{max5}
\end{align}
for a DMC $W$ such that $|\Pi_{W, \Gamma}^*| = 1$ and $V(\Gamma) > 0$, where the function $\mathcal{K} : \mathbb{R} \times (0, \infty) \to (0, 1)$ is given by  
\begin{align}
    \mathcal{K}\left(r, V \right) &= \min_{\substack{\Pi:\\
    \mathbb{E}[\Pi] = r \\
    \text{Var}(\Pi) \leq V\\
    |\text{supp}(\Pi)| \leq 3
    }} \mathbb{E}\left [\Phi(\Pi) \right]. \label{min5} 
\end{align}
The maximum and the minimum in $(\ref{max5})$ and $(\ref{min5})$, respectively, are attained \cite[Lemmas  3 and 4]{mahmood2024channel}. 

It is worth describing here the code construction that achieves the optimal SOCR in $(\ref{max5})$. From \cite{mahmood2024channelcodingmeanvariance}, the achievability coding scheme makes use of the solution to the optimization problem in 
$$\mathcal{K}\left(\frac{r^*}{\sqrt{V(\Gamma) }}, \frac{C'(\Gamma)^2 V}{V(\Gamma)} \right),$$
where $r^* = r_{\text{m.v.}}(\epsilon, \Gamma, V) $. Specifically, the minimizing probability distribution with three point masses, denoted by 
\begin{align*}
    P_{\Pi}(\pi) = \begin{cases}
    p_1 & \pi = \pi_1\\
    p_2 & \pi = \pi_2\\
    p_3 & \pi = \pi_3,
    \end{cases}
\end{align*}
is mapped to three different cost values $\Gamma_1, \Gamma_2$ and $\Gamma_3$ as follows:
$$\Gamma_{j} = \Gamma - \frac{\sqrt{V(\Gamma) }}{C'(\Gamma)\sqrt{n}}\pi_j + \frac{r^*}{C'(\Gamma)\sqrt{n}}$$
for $j \in \{1,2,3 \}$. The cost values $(\Gamma_1,\Gamma_2,\Gamma_3)$ are in turn mapped to three types $T_1, T_2$ and $T_3$, each type $T_j$ being close to a capacity-cost-achieving distribution for cost $\Gamma_j$. Subsequently, a random coding scheme is used where the codewords are drawn randomly from one of the three type classes with probability weights $p_1, p_2$ and $p_3$. In contrast, for the a.s. constraint, the random coding scheme usually involves drawing codewords from only one type class.

\begin{theorem}
Fix an arbitrary $\epsilon \in (0, 1)$. Then for any $\Gamma \in (\Gamma_0, \Gamma^*)$, $V > 0$ and a DMC $W$ such that $|\Pi_{W, \Gamma}^*| = 1$ and $V(\Gamma) > 0$, we have $r_{\text{m.v.}}(\epsilon, \Gamma, V) > r_{\text{a.s.}}(\epsilon, \Gamma)$.
    \label{thm2}
\end{theorem}
\textit{Proof:} The proof is given in Section \ref{thm2proof}. 

\begin{figure}[H]
    \centering
\includegraphics[width=8.8cm]{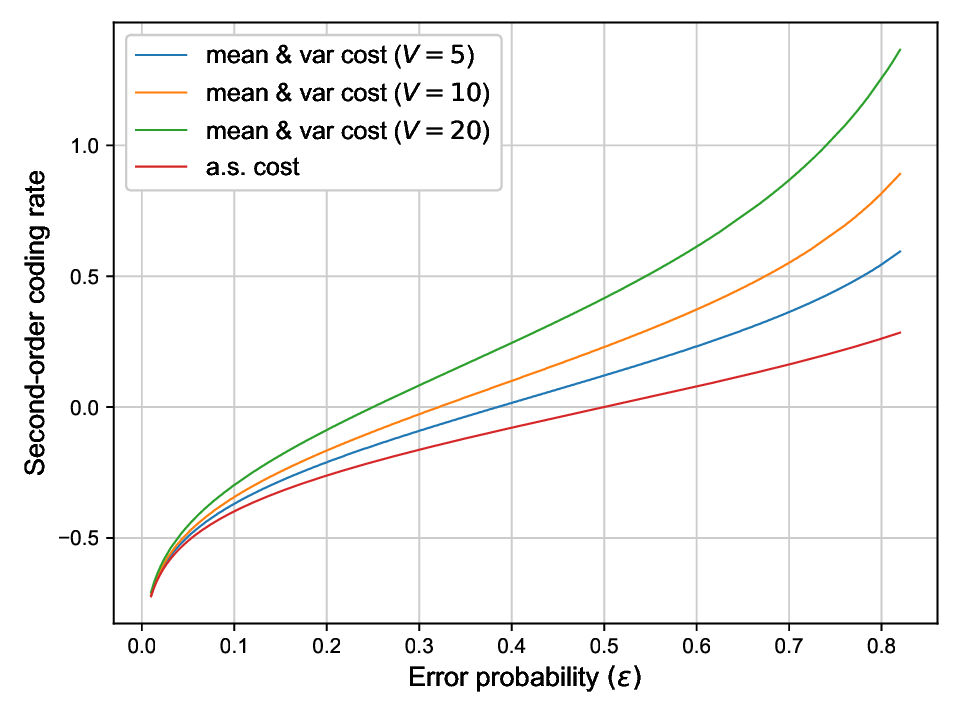}
\caption{The SOCR is compared between the almost-sure cost constraint and the mean and variance cost constraints for different values of $V$. The plots for the mean and variance cost constraints are lower bounds to the SOCR since they are obtained through a non-exhaustive search of the feasible region in the maximization and minimization in $(\ref{max5})$ and $(\ref{min5})$, respectively. Insofar as the lower bounds are close to the optimal SOCR, the plots suggest that for a fixed $V$, the improvement over the a.s. cost constraint diminishes as $\epsilon \to 0$.}
\label{bsc}
\end{figure}

\begin{figure}[H]
    \centering
\includegraphics[width=8.8cm]{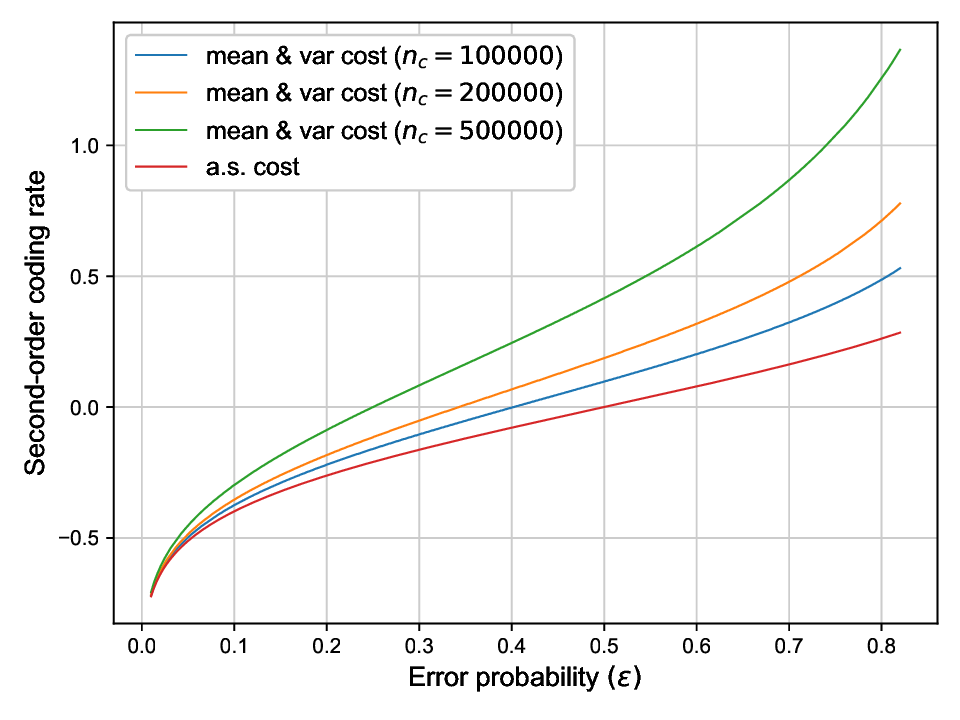}
\caption{The SOCR is compared between the almost-sure cost constraint and the mean and variance cost constraints for an $(\alpha, \beta)$-ergodic random codebook with $\alpha = \beta = 0.1$.}
\label{bsc2}
\end{figure}

The improvement in Theorem \ref{thm2} is shown in Fig. \ref{bsc} for a binary symmetric channel. Specifically, the second-order coding rate as a function of the average error probability is shown in Fig. \ref{bsc} and Fig. \ref{bsc2} for a binary symmetric channel with parameter $p = 0.3$, alphabets $\mathcal{A} = \mathcal{B} = \{0, 1 \}$, cost threshold $\Gamma = 0.2$ and cost function $c(x) = x$.   

As discussed in $(\ref{criticaln})$, the choice of $V$ together with the desired values of $\alpha$ and $\beta$ specifies the critical blocklength exceeding which guarantees the $(\alpha, \beta)$-ergodicity of the coding scheme. In practice, the choice of blocklength is more fundamental as it affects complexity and latency. Therefore, it is more prudent for the value of $V$ to be dictated by the blocklength $n$ and the desired $(\alpha, \beta)$-ergodicity via the relation $V = n \beta \alpha^2 \Gamma^2$ derived from $(\ref{criticaln})$. With the same channel $(p=0.3)$ and cost $(\Gamma = 0.2)$ parameters as used in Fig. 1, Fig. \ref{bsc2} shows the second-order performance for different critical blocklengths for an $(\alpha, \beta)$-ergodic codebook with $\alpha = \beta = 0.1$. 

\subsection{Performance improvement for feedback codes}

\begin{definition}
    A controller is a function $F : (\mathcal{A} \times \mathcal{B})^* \to \mathcal{P}(\mathcal{A})$. 
\end{definition}

Random feedback codes can be constructed by controllers. Given a message $m \in \mathcal{M}_R$ and the past channel inputs and outputs $(x^{k-1}, y^{k-1})$, the channel input $X_k = f(m, x^{k-1}, y^{k-1})$ at time instant $k$ is distributed according to $F(x^{k-1}, y^{k-1})$. There is a one-to-one correspondence between a random feedback code and a controller-based code. 
\begin{itemize}
    \item A random feedback code $(f, g)$ is equivalently specified by the joint distribution  
\begin{align}
    &p_{M, X^n, Y^n, \hat{M}} = \mbox{} \notag \\
    &\quad \quad p_M \left( \prod_{i=1}^n p_{X_i|M, X^{i-1}, Y^{i-1}} p_{Y_i|X_i} \right) p_{\hat{M}| Y^n},      \notag 
\end{align}
where $\hat{M}$ is the decoded message. Marginalizing out $M$ and $\hat{M}$, we obtain 
\begin{align*}
    p_{X^n, Y^n} = \prod_{i=1}^n p_{X_i|X^{i-1}, Y^{i-1}}\, p_{Y_i|X_i}
\end{align*}
from which a controller $F$ can be obtained with the specification 
\begin{align}
    F(x_k|x^{k-1}, y^{k-1}) = p_{X_k|X^{k-1}, Y^{k-1}}(x_k|x^{k-1}, y^{k-1})   \notag 
\end{align}
for each time $k$. 
\item Likewise, a controller $F$ specifies a random feedback code by inducing the following joint distribution: 
\begin{align*}
    &p_{M, X^n, Y^n, \hat{M}}(m, x^n, y^n, \hat{m}) \\
    &= p_M(m) \left( \prod_{i=1}^n F(x_i|x^{i-1}, y^{i-1}) p_{Y_i|X_i}(y_i|x_i) \right) \cdot \mbox{}\\
    & \quad \quad \quad \quad \quad \quad \quad \quad \quad \quad \quad p_{\hat{M}| Y^n}(\hat{m}|y^n).       
\end{align*}
\end{itemize}

A controller $F$ along with the channel $W$ specify a joint distribution over $\mathcal{A}^n \times \mathcal{B}^n$ with probability assignments given by 
\begin{align}
    (F \circ W)(x^n, y^n) = \prod_{k=1}^n F(x_k|x^{k-1}, y^{k-1}) W(y_k|x_k). \label{probcons}
\end{align}

\begin{lemma}
Consider a channel $W$ with cost constraint $\Gamma \in (\Gamma_0, \Gamma^*)$. Then for every $n, \rho > 0$ and $\epsilon \in (0, 1)$,   
\begin{align}
   &\log M^*_{\text{fb}}(n, \epsilon,  \Gamma) \leq \log \rho - \log \Bigg [ \Bigg ( 1 - \epsilon - \mbox{} \notag \\
   & \quad \quad \quad \sup_{F}\, \inf_{q \in \mathcal{P}(\mathcal{B}^n)} (F \circ W) \left( \frac{W(Y^n|X^n)}{q(Y^n)} > \rho \right)\Bigg )^+\Bigg ], 
   \label{qandp}
\end{align}
where the supremum in $(\ref{qandp})$ is over all controllers $F$ satisfying 
\begin{align}
    &(F \circ W)\left( c(X^n) \leq \Gamma \right) \notag \\
    &=  \sum_{x^n: c(x^n)\leq \Gamma} \sum_{y^n \in \mathcal{B}^n} \prod_{k=1}^n F(x_k|x^{k-1}, y^{k-1}) W(y_k|x_k) \notag  \\
    &\,= 1. \label{vc}
\end{align}
\label{qandp2}
\end{lemma}

\textit{Proof:} Lemma \ref{qandp2} is similar to \cite[Theorem 27]{5452208}, \cite[(42)]{8012458}, \cite[Lemma 15]{9099482}, \cite[Lemma 2]{mahmood2024channel} and others. Its proof is omitted.

\begin{theorem}
Fix an arbitrary $\epsilon \in (0, 1)$. Then for any $\Gamma \in (\Gamma_0, \Gamma^*)$ and a simple-dispersion DMC $W$ such that $V(\Gamma) > 0$, we have $r_{\text{a.s.,fb}}(\epsilon, \Gamma) = r_{\text{a.s.}}(\epsilon, \Gamma)$.
\label{thm}
\end{theorem}

\textit{Proof:} The proof is given in Section \ref{thm_proof}.

We only prove a converse result that is the following upper bound: 
    \begin{align}
       r_{\text{a.s.,fb}}(\epsilon, \Gamma) \leq \sqrt{V(\Gamma)} \Phi^{-1}(\epsilon). \label{4x}  
    \end{align}
    The result of Theorem \ref{thm} follows by combining $(\ref{4x})$ with the existing achievability result (without feedback) from \cite[Theorem 3]{5290292}.

From Theorems \ref{thm2} and \ref{thm} alone, we have that $r_{\text{m.v.,fb}}(\epsilon, \Gamma, V) > r_{\text{a.s.,fb}}(\epsilon, \Gamma, V)$. More importantly, the mean and variance cost formulation admits feedback mechanisms that improve the SOCR, even if the capacity-cost-achieving distribution is unique, i.e., $|\Pi_{W, \Gamma}^*| = 1$. This is the more interesting case since for compound-dispersion channels where $|\Pi_{W, \Gamma}^*| > 1$, feedback is already known to improve second-order performance via timid/bold coding \cite{9099482}.

In contrast to the almost-sure constraint where $r_{\text{a.s.,fb}}(\epsilon, \Gamma) = r_{\text{a.s.}}(\epsilon, \Gamma)$, we observe that $r_{\text{m.v.,fb}}(\epsilon, \Gamma, V) > r_{\text{m.v.}}(\epsilon, \Gamma, V)$ for simple-dispersion channels with $|\Pi_{W, \Gamma}^*| = 1$ \cite[Theorem 3]{mahmood2024channel}. In summary, for any $\epsilon \in (0, 1)$, $\Gamma \in (\Gamma_0, \Gamma^*)$, $V > 0$ and a simple-dispersion DMC $W$ such that $|\Pi_{W, \Gamma}^*| = 1$ and $V(\Gamma) > 0$,  
\begin{align*}
    r_{\text{m.v.,fb}}(\epsilon, \Gamma, V) > r_{\text{m.v.}}(\epsilon, \Gamma, V) > r_{\text{a.s.,fb}}(\epsilon, \Gamma) = r_{\text{a.s.}}(\epsilon, \Gamma),    
\end{align*}
where the last equality above has been proven without the assumption $|\Pi_{W, \Gamma}^*| = 1$. 
  
\section{Proof of Theorem \ref{thm2} \label{thm2proof}}

Since $\mathcal{K}(r, V)$ is a continuous function \cite[Lemma 3]{mahmood2024channel}, it suffices to show that for all $r \in \mathbb{R}$ and $V > 0$,  
\begin{align}
    \min_{\substack{\Pi:\\
    \mathbb{E}[\Pi] = r \\
    \text{Var}(\Pi) \leq V\\
    |\text{supp}(\Pi)| \leq 2
    }} \mathbb{E}\left [\Phi(\Pi) \right] < \Phi(r). \label{bv}
\end{align}

The LHS of $(\ref{bv})$ can be written as 
\begin{align*}
    \min_{\substack{p, \pi:\\
    0 \leq p \leq 1\\
    \frac{p}{1-p}(\pi - r)^2 \leq V}} \left [ p \Phi(\pi) + (1-p) \Phi\left( \frac{r - p \pi}{1-p} \right) \right ],  
\end{align*}
where we used the constraint $\mathbb{E}[\Pi] = r$ to eliminate one of the decision variables.

Assume by contradiction that 
\begin{align}
    p \Phi(\pi) + (1-p) \Phi\left( \frac{r - p \pi}{1-p} \right) \geq \Phi(r) \label{eq}
\end{align}
for all $\pi \geq r$, $p \in [0, 1]$ and $\frac{p}{1-p}(\pi - r)^2 \leq V$. The assumption $\pi \geq r$ is without loss of generality since one of the two point masses must be greater than or equal to $r$. 

If $(\ref{eq})$ holds, then 
\begin{align*}
    p \Phi(\pi) + (1-p) \Phi\left( \frac{r - p \pi}{1-p} \right) \geq \Phi(r)
\end{align*}
for all $\pi \geq r$, $p \in [0, 1]$ and $\frac{p}{1-p}(\pi - r)^2 = V$. Since $\pi = r + \sqrt{\frac{V(1-p)}{p}}$ in this case, we must have 
\begin{align}
    &p\, \Phi\left( r + \sqrt{\frac{V(1-p)}{p}} \right ) + \mbox{} \notag  \\
    & \quad   (1-p) \Phi\left( \frac{r}{1-p} - \frac{p}{1-p}\left( r + \sqrt{\frac{V(1-p)}{p}}\right) \right) \geq \Phi(r) \notag \\
    &p\, \Phi\left( r + \sqrt{\frac{V(1-p)}{p}} \right ) + \mbox{} \notag \\
    & \quad \quad \quad \quad  (1-p) \Phi\left( r  - \frac{p}{1-p}  \sqrt{\frac{V(1-p)}{p}}\right)  \geq \Phi(r) \notag \\
    &p\, \Phi\left( r + \sqrt{\frac{V(1-p)}{p}} \right ) + (1-p) \Phi\left( r  -   \sqrt{\frac{V p}{1-p}}\right)  \geq \Phi(r) \label{gp}
\end{align}
for all $p \in [0, 1]$. 

Consider the function 
\begin{align*}
    &f(p) = p\, \Phi\left( r + \sqrt{\frac{V(1-p)}{p}} \right ) + \mbox{} \\
    & \quad \quad \quad \quad  (1-p) \Phi\left( r  -   \sqrt{\frac{V p}{1-p}}\right)  - \Phi(r)
\end{align*}
with domain $p \in [0, 1]$. For any $p \in (0, 1)$,  
\begin{align}
    &\frac{f(p) - f(0)}{p} \notag  \\
    &=  \Phi\left( r + \sqrt{\frac{V(1-p)}{p}} \right ) + \frac{1-p}{p} \Phi\left( r  -   \sqrt{\frac{V p}{1-p}}\right)  - \frac{\Phi(r)}{p} \notag \\
    &\leq \Phi\left( r + \sqrt{\frac{V(1-p)}{p}} \right ) + \frac{1}{p} \Phi\left( r  -   \sqrt{\frac{V p}{1-p}}\right)  - \frac{\Phi(r)}{p} \notag \\
    &\stackrel{(a)}{=} \Phi\left( r + \sqrt{\frac{V(1-p)}{p}} \right ) + \frac{1}{p} \left [ \Phi(r) - \phi(\tilde{r}) \sqrt{\frac{V p}{1-p}} - \Phi(r)  \right ] \notag \\
    &= \Phi\left( r + \sqrt{\frac{V(1-p)}{p}} \right ) -    \phi(\tilde{r}) \sqrt{\frac{V }{p(1-p)}}. \label{co}
\end{align}
In equality $(a)$, we have $r - \sqrt{\frac{V p}{1-p}} < \tilde{r} < r$ by the mean value theorem. It is easy to see that for sufficiently small $p > 0$, the expression in $(\ref{co})$ is negative. Since $f(0) = 0$, we have $f(p) < 0$ for some $p > 0$, which contradicts $(\ref{gp})$.

\section{Proof of Theorem \ref{thm} \label{thm_proof}}

For any $t \in \mathcal{P}_n(\mathcal{A})$, define  
\begin{align*}
    d_W(t) := \inf_{P \in \Pi_{W,\Gamma}^*} ||t - P||_1. 
\end{align*}

For any $0 < \gamma \leq \frac{V(\Gamma)}{4|\mathcal{A}| \nu_{\max}}$, define  
\begin{align}
\begin{split}
    \mathcal{P}_{n}^\gamma &= \left \{ x^n \in \mathcal{A}^n: c(x^n) \leq \Gamma \text{ and } d_W(t(x^n)) > \gamma  \right \}\\
    \mathcal{P}_{n}^{\gamma, c} &= \left \{ x^n \in \mathcal{A}^n: c(x^n) \leq \Gamma \text{ and } d_W(t(x^n)) \leq \gamma  \right \}.
\end{split}
\label{splitsets}
\end{align}

\begin{definition}
    For any distribution $P \in \mathcal{P}(\mathcal{A})$ and $S \subset \mathcal{A}$ such that $P(S) > 0$, define the probability measure 
    \begin{align*}
        P |_S(x) = \begin{cases}
            \frac{P(x)}{P(S)} & x \in S\\
            0 & \text{otherwise}.
        \end{cases} 
    \end{align*}
\end{definition}

\begin{definition}
    For any $k \geq 0$ and any $x^k \in \mathcal{A}^k$, let\footnote{For $k \geq n$, $\mathcal{A}_{x^k} = \emptyset$ and for $k=0$, $(x^k, x) = (x,)$.}
\begin{align*}
    \mathcal{A}_{x^k} = \left \{ x \in \mathcal{A}: (x^k, x) \text{ is a prefix of some } x^n \in \mathcal{P}_{n}^{\gamma, c}  \right \}.
\end{align*}
Fix $ (a_0, a_1, \ldots, a_{n-1}) \in \mathcal{P}_{n}^{\gamma, c}$ arbitrarily and let $\mathds{1}_{a_i} \in \mathcal{P}(\mathcal{A})$ denote a single point-mass distribution at $a_i$. Then for any $0 \leq k \leq n-1$, $x^k \in \mathcal{A}^k, y^k \in \mathcal{B}^k$ and a controller $F$ satisfying $(\ref{vc})$, we define the controller $F_\gamma$ as  
\begin{align*}
    &F_\gamma(x^k, y^k) := \mbox{} \\
    &\quad  \begin{cases}
        F(x^k, y^k)|_{\mathcal{A}_{x^k}} & \text{ if } F( \mathcal{A}_{x^k} | x^k, y^k) > 0\\
        \text{Unif}(A_{x^k}) & \text{ if } F( \mathcal{A}_{x^k} | x^k, y^k) = 0  \text{ and } | \mathcal{A}_{x^k}| \neq 0\\
        \mathds{1}_{a_k} & \text{ otherwise}.
    \end{cases}  
\end{align*}
\label{w}
\end{definition}
\begin{remark}
    Given any controller $F$ satisfying $(\ref{vc})$, Definition $\ref{w}$ constructs a modified controller $F_\gamma$ which satisfies 
\begin{align}
    &(F_\gamma \circ W) \left( \mathcal{P}_{n}^{\gamma, c}\right) \notag  \\
    &\quad \quad := \sum_{x^n \in \mathcal{P}_n^{\gamma, c}} \sum_{y^n \in \mathcal{B}^n} \prod_{k=1}^n F_\gamma(x_k|x^{k-1}, y^{k-1}) W(y_k|x_k) \notag  \\
    &\quad \quad= 1. \label{eq:modcost}
\end{align}
Intuitively, $F_\gamma$ amplifies the probability assignments of $F$ over the set $\mathcal{P}_{n}^{\gamma, c}$ and nullifies the probability assignments of $F$ over the set $\mathcal{P}_{n}^{\gamma}$ so that $X^n \in \mathcal{P}_{n}^{\gamma, c}$ almost surely for $(X^n, Y^n) \sim (F_\gamma \circ W)$. 
Definition~\ref{w} is inspired by, and corrects an error in,~\cite[Def.~8]{9099482}.
With the definition given in \cite[Def.~8]{9099482}, 
the analogue of (\ref{eq:modcost}) does not hold, although it is asserted in
the proof of \cite[Thm.~3]{9099482}. This can be rectified by using Definition~\ref{w}
in place of \cite[Def.~8]{9099482}. An analogous comment applies to the next definition and \cite[Def.~9]{9099482}.
\label{ha}
\end{remark}

\begin{definition}
    For any type $t \in \mathcal{P}_n(\mathcal{A})$ such that $T^n_{\mathcal{A}}(t) \subset \mathcal{P}_n^\gamma$, $k \geq 0$ and any $x^k \in \mathcal{A}^k$, let
\begin{align*}
    \mathcal{A}_{x^k}^t = \left \{ x \in \mathcal{A}: (x^k, x) \text{ is a prefix of some } x^n \in T^n_{\mathcal{A}}(t)  \right \}.
\end{align*}
Fix $(a_0, a_1, \ldots, a_{n-1}) \in T^n_{\mathcal{A}}(t)$ arbitrarily and let $\mathds{1}_{a_i} \in \mathcal{P}(\mathcal{A})$ denote a single point-mass distribution at $a_i$. Then for any $0 \leq k \leq n-1$, $x^k \in \mathcal{A}^k, y^k \in \mathcal{B}^k$ and a controller $F$ satisfying $(\ref{vc})$, we define the controller $F_t$ as  
\begin{align*}
    &F_t(x^k, y^k):= \\
    & \begin{cases}
        F(x^k, y^k)|_{\mathcal{A}_{x^k}^t} & \text{ if } F( \mathcal{A}^t_{x^k} | x^k, y^k) > 0\\
        \text{Unif}(A^t_{x^k}) & \text{ if } F( \mathcal{A}^t_{x^k} | x^k, y^k) = 0  \text{ and } | \mathcal{A}_{x^k}| \neq 0\\
        \mathds{1}_{a_k} & \text{ otherwise}.
    \end{cases}  
\end{align*}
\label{w2}
\end{definition}

\begin{remark}
\label{remarktype}
    Given any controller $F$ satisfying $(\ref{vc})$, Definition $\ref{w2}$ constructs a modified controller $F_t$ which satisfies $(F_t \circ W) \left(T^n_{\mathcal{A}}(t)\right) = 1$ for $T^n_{\mathcal{A}}(t) \subset \mathcal{P}_n^\gamma$. 
\end{remark}

Now let\footnote{The choice of the output distribution in $(\ref{choiceq})$ is similar to what has appeared in previous works, e.g., \cite[(37)]{8012458} and \cite[(87)]{9099482}.}  
\begin{align}
    q(y^n) = \frac{1}{2} \prod_{i=1}^n Q^*(y_i) + \frac{1}{2} \frac{1}{|\mathcal{P}_n(\mathcal{A})|} \sum_{t \in \mathcal{P}_n(\mathcal{A})} \prod_{i=1}^n q_t(y_i), \label{choiceq}
\end{align}
where
\begin{align*}
    q_t(b) := \sum_{a \in \mathcal{A}} t(a) W(b|a).      
\end{align*} 
Let $P$ denote the distribution $F \circ W$. Let $P_\gamma$ denote the distribution $F_\gamma \circ W$. Let $P_t$ denote the distribution $F_t \circ W$ for each $t \in \mathcal{P}_n(\mathcal{A})$ such that $T^n_{\mathcal{A}}(t) \subset \mathcal{P}_n^\gamma$. Note that all controllers $F, F_\gamma$ and $F_t$ satisfy $(\ref{vc})$. We have 
\begin{align}
    & P\left( \frac{W(Y^n|X^n)}{q(Y^n)} > \rho \right) \notag \\
    &= P\left( \frac{W(Y^n|X^n)}{q(Y^n)} > \rho \cap d_W(t(X^n)) \leq \gamma \right) + \mbox{} \notag \\
    & \quad \quad \quad \quad \quad \quad P\left( \frac{W(Y^n|X^n)}{q(Y^n)} > \rho \cap d_W(t(X^n)) > \gamma \right) \notag  \\
    &= P\left( \frac{W(Y^n|X^n)}{q(Y^n)} > \rho \cap d_W(t(X^n)) \leq \gamma \right) + \mbox{} \notag \\
    & \quad \quad \quad \sum_{t: T^n_{\mathcal{A}}(t) \subset \mathcal{P}_n^\gamma } P\left( \frac{W(Y^n|X^n)}{q(Y^n)} > \rho \cap t(X^n) = t \right) \notag \\
    &\stackrel{(a)}{\leq} P_\gamma\left( \frac{W(Y^n|X^n)}{q(Y^n)} > \rho  \right) + \mbox{} \notag \\
    & \quad \quad \quad \quad \quad \quad \sum_{t: T^n_{\mathcal{A}}(t) \subset \mathcal{P}_n^\gamma} P_t\left( \frac{W(Y^n|X^n)}{q(Y^n)} > \rho  \right). \label{q}
\end{align}
Inequality $(a)$ follows from the following argument. For any $(x^n, y^n)$ such that $x^n \in \mathcal{P}_{n}^{\gamma, c}$, note that for all $1 \leq k \leq n$, $\mathcal{A}_{x^{k-1}} \neq \emptyset$ and $x_{k} \in \mathcal{A}_{x^{k-1}}$ so that 
\begin{align}
    F(x_k|x^{k-1}, y^{k-1}) \leq F(\mathcal{A}_{x^{k-1}} | x^{k-1}, y^{k-1}). \label{fw}
\end{align} 
Then   
\begin{align*}
    &P_\gamma\left( (x^n, y^n) \right)\\
    &= \prod_{k=1}^n F_\gamma(x_k|x^{k-1}, y^{k-1}) W(y_k|x_k)\\
    &\stackrel{(b)}{\geq} \prod_{k=1}^n \frac{F(x_k|x^{k-1}, y^{k-1})}{F(\mathcal{A}_{x^{k-1}} | x^{k-1}, y^{k-1})} W(y_k|x_k)\\
    &\geq \prod_{k=1}^n F(x_k|x^{k-1}, y^{k-1}) W(y_k|x_k)\\
    &= P((x^n, y^n)).
\end{align*}
With some abuse of notation, we assume in inequality $(b)$ above that if $F(\mathcal{A}_{x^{k-1}} | x^{k-1}, y^{k-1}) = 0$, then 
$$\frac{F(x_k|x^{k-1}, y^{k-1})}{F(\mathcal{A}_{x^{k-1}} | x^{k-1}, y^{k-1})} = 0$$
which is justified by $(\ref{fw})$. 

A similar derivation gives 
\begin{align*}
    P_t((x^n, y^n)) \geq P(x^n, y^n)
\end{align*}
for all $(x^n, y^n)$ such that $c(x^n) \leq \Gamma$ and $t(x^n) =t$, where $T^n_{\mathcal{A}}(t) \subset \mathcal{P}_n^\gamma$. 

Let $\rho = \exp\left(n C(\Gamma) + \sqrt{n} r \right)$, where $r$ will be specified later. Define\footnote{This proof follows that
    of~\cite[Thm.~3]{9099482} and corrects an error therein: the $\mathcal{F}_i$
filtration defined before (98) should be defined like $\mathcal{G}_i$ here.}
\begin{align*}
    \mathcal{G}_i &:= \sigma(X_1, \ldots, X_{i+1}, Y_1, \ldots, Y_i)\\
    Z_i &:=   i(X_i,Y_i)- \mathbb{E}\left [i(X_i, Y_i)|\mathcal{G}_{i-1} \right]\\
    \mathcal{F}_i &:= \sigma(Z_1, Z_2, \ldots, Z_{i}).
\end{align*}
Two things are important to note here. First, 
by the Markov property $(X^{i-1}, Y^{i-1}) - X_i - Y_i$, 
we have $\mathbb{E}\left [i(X_i, Y_i)|\mathcal{G}_{i-1} \right] = \mathbb{E}\left [i(X_i, Y_i)|X_i \right]$ a.s. Second, $\mathcal{F}_i \subset \mathcal{G}_i$.

For the first term in $(\ref{q})$, we can upper bound it as follows: 
\begin{align}
    &P_\gamma\left( \frac{W(Y^n|X^n)}{q(Y^n)} > \rho  \right) \notag \\
    &\leq P_\gamma\left( \prod_{i=1}^n \frac{W(Y_i|X_i)}{Q^*(Y_i)} > \frac{\rho}{2}  \right) \notag \\
    &= P_\gamma\left( \sum_{i=1}^n \left [  \log\left( \frac{W(Y_i|X_i)}{Q^*(Y_i)} \right)- C(\Gamma) \right ] > \sqrt{n} r - \log(2)   \right) \notag  \\
    &\stackrel{(a)}{\leq} P_\gamma\left( \sum_{i=1}^n \left [  i(X_i,Y_i)- \mathbb{E}\left [i(X_i, Y_i)|\mathcal{G}_{i-1} \right]  \right ] > \sqrt{n} r - \log(2)   \right) \notag  \\
    &= P_\gamma\left( \sum_{i=1}^n Z_i > \sqrt{n} r - \log(2)   \right). \label{MCLT}
\end{align}
In inequality $(a)$, we used the following lemma and the fact that $c(X^n) \leq \Gamma$ almost surely.

\begin{lemma}
For $\Gamma \in (\Gamma_0, \Gamma^*)$,  
    \begin{align}
    \mathbb{E}\left [ i(X, Y) | X \right ] \leq C(\Gamma) - C'(\Gamma) \left( \Gamma - c(X) \right)
    \end{align}
    almost surely, where $X$ has an arbitrary distribution and $Y$ is the output of the channel $W$ when $X$ is the input. 
    \label{getout}
\end{lemma}

\textit{Proof:} See \cite[Proposition 1]{mahmood2024channelcodingmeanvariance} and its references. 

We will now apply a martingale central limit theorem \cite[Corollary to Theorem 2]{Bolthausen1982} to the expression in $(\ref{MCLT})$. We first verify that the hypotheses of \cite[Corollary to Theorem 2]{Bolthausen1982} are satisfied:
\begin{enumerate}
    \item First, we require that 
    $$\max_{1 \leq k \leq n} |Z_k| < \infty.$$
    Since $Q^*(b) > 0$ for all $b \in \mathcal{B}$ by assumption and $W(Y_k|X_k) > 0$ almost surely for each channel input and output pair $(X_k, Y_k)$, we have 
\begin{align*}
    |Z_k| &\leq  \max_{a \in \mathcal{A}, b \in \mathcal{B} : W(b|a) > 0} 2\,| i(a, b)  |\\
    &:= 2 i_{\max} < \infty
\end{align*}
for all $1\leq k \leq n$. 
\item Second, we require that 
\begin{align*}
    \mathbb{E}\left [Z_k| \mathcal{F}_{k-1} \right ] = 0
\end{align*}
almost surely for all $1 \leq k \leq n$ \cite[p. 672]{Bolthausen1982}. This is true because $\mathbb{E}\left [Z_k| \mathcal{G}_{k-1} \right ] = 0$ implies  
\begin{align*}
    \mathbb{E}\left [ \mathbb{E}\left [Z_k| \mathcal{G}_{k-1} \right ] | \mathcal{F}_{k-1}  \right] &= 0\\
    \mathbb{E}\left [Z_k| \mathcal{F}_{k-1} \right ] &= 0.
\end{align*}
 
\end{enumerate}

Under the above two conditions, it follows from \cite[Corollary to Theorem 2]{Bolthausen1982} that there exists a constant $\kappa > 0$ depending only on $i_{\max}$ such that for any $s \in \mathbb{R}$,
\begin{align}
    &P_\gamma\left( \frac{1}{\sqrt{\sum_{i=1}^n \mathbb{E}\left [ Z_i^2 \right ] }} \sum_{i=1}^n Z_i \leq s  \right) \geq \Phi\left( s\right) - \mbox{} \notag \\
    &\kappa \left [ \frac{n \log n}{\left(\sum_{i=1}^n \mathbb{E}_\gamma\left [ Z_i^2 \right ] \right)^{\frac{3}{2}}} + \Bigg | \Bigg | \frac{\sum_{i=1}^n \mathbb{E}_\gamma[Z_i^2| \mathcal{F}_{i-1}]}{\sum_{i=1}^n \mathbb{E}_\gamma[Z
    _i^2]} - 1  \Bigg | \Bigg |_\infty^{1/2} \right ]. \label{conv}
\end{align}
Using Lemma \ref{restatedlemma} in $(\ref{conv})$, we obtain 
\begin{align}
    &P_\gamma\left( \frac{1}{\sqrt{\sum_{i=1}^n \mathbb{E}\left [ Z_i^2 \right ] }} \sum_{i=1}^n Z_i \leq s  \right) \geq \Phi\left( s\right) - \mbox{} \notag \\
    &\kappa \left [ \frac{n \log n}{\left(n V(\Gamma) - n|\mathcal{A}| \gamma \nu_{\max} \right)^{\frac{3}{2}}} + \left(\frac{ 4|\mathcal{A}| \gamma \nu_{\max}}{V(\Gamma)} \right)^{1/2} \right ] \notag \\
    & \quad \quad \quad \quad \quad \quad \quad \quad  \geq \Phi(s) - \beta_\gamma, \label{finres} 
\end{align}
where the last inequality holds for sufficiently large $n$ for some constant $\beta_\gamma > 0$ which can be chosen such that $\beta_\gamma \to 0$ as $\gamma \to 0$. 

\begin{lemma}
\label{restatedlemma}
We have 
   \begin{align*}
     &V(\Gamma) - 2 \gamma \nu_{\max}  \leq \frac{1}{n}\sum_{i=1}^n \mathbb{E}_\gamma\left [ Z_i^2 \right] \leq  V(\Gamma) + 2 \gamma \nu_{\max}.
     \end{align*}
     Furthermore, for $\gamma \leq \frac{V(\Gamma)}{4 \nu_{\max}}$, 
     \begin{align*}
     &\Bigg | \Bigg | \frac{\sum_{i=1}^n \mathbb{E}_\gamma[Z_i^2| \mathcal{F}_{i-1}]}{\sum_{i=1}^n \mathbb{E}_\gamma[Z
    _i^2]} - 1  \Bigg | \Bigg |_\infty \leq \frac{ 8 \gamma \nu_{\max}}{V(\Gamma)}
\end{align*}
almost surely according to the probability measure $P_\gamma$. 
\end{lemma}
\textit{Proof}: The proof of Lemma \ref{restatedlemma} is given in Appendix \ref{restatedlemma_proof}.

Using the result in $(\ref{finres})$ and Lemma \ref{restatedlemma} in the expression in $(\ref{MCLT})$, we obtain 
\begin{align}
    &P_\gamma\left( \sum_{i=1}^n Z_i > \sqrt{n} r - \log(2)   \right) \notag \\
    &= P_\gamma\left( \frac{1}{\sqrt{\sum_{i=1}^n \mathbb{E}_\gamma\left [ Z_i^2 \right]}} \sum_{i=1}^n Z_i >  \frac{ \sqrt{n} r - \log(2)}{\sqrt{\sum_{i=1}^n \mathbb{E}_\gamma\left [ Z_i^2 \right]}}   \right) \notag \\
    &= 1 - P_\gamma\left( \frac{1}{\sqrt{\sum_{i=1}^n \mathbb{E}_\gamma\left [ Z_i^2 \right]}} \sum_{i=1}^n Z_i \leq \frac{ \sqrt{n} r - \log(2)}{\sqrt{\sum_{i=1}^n \mathbb{E}_\gamma\left [ Z_i^2 \right]}}   \right) \notag \\
    &\leq \begin{cases}
        1 - P_\gamma\left( \frac{1}{\sqrt{\sum_{i=1}^n \mathbb{E}_\gamma\left [ Z_i^2 \right]}} \sum_{i=1}^n Z_i \leq  \frac{  r - \frac{\log(2)}{\sqrt{n}}}{\sqrt{ V(\Gamma) + |\mathcal{A}| \gamma \nu_{\max}}}   \right) & \\
\quad\quad\quad\quad\quad\quad\quad\quad\quad\quad\quad\quad \text{ if } r \geq \frac{\log 2}{\sqrt{n}} \\
        1 - P_\gamma\left( \frac{1}{\sqrt{\sum_{i=1}^n \mathbb{E}_\gamma\left [ Z_i^2 \right]}} \sum_{i=1}^n Z_i \leq  \frac{  r - \frac{\log(2)}{\sqrt{n}}}{\sqrt{ V(\Gamma) - |\mathcal{A}| \gamma \nu_{\max}}}   \right) & \\
\quad\quad\quad\quad\quad\quad\quad\quad\quad\quad\quad\quad\text{ if } r < \frac{\log 2}{\sqrt{n}}
    \end{cases} \notag \\
    &\leq \begin{cases}
        1 - \Phi\left( \frac{  r - \frac{\log(2)}{\sqrt{n}}}{\sqrt{ V(\Gamma) + |\mathcal{A}| \gamma \nu_{\max}}}   \right) + \beta_\gamma & \text{ if } r \geq \frac{\log 2}{\sqrt{n}}\\
        1 - \Phi\left(  \frac{  r - \frac{\log(2)}{\sqrt{n}}}{\sqrt{ V(\Gamma) - |\mathcal{A}| \gamma \nu_{\max}}}   \right) + \beta_\gamma & \text{ if } r < \frac{\log 2}{\sqrt{n}}.
    \end{cases} \label{userhere}
\end{align}
Let 
\begin{align}
    r = \begin{cases}
        \sqrt{V(\Gamma) + |\mathcal{A}| \gamma \nu_{\max} }\, \Phi^{-1}\left(\epsilon + 3 \beta_\gamma  \right) + \frac{\log 2}{\sqrt{n}}  & \\
        \quad \quad \quad \quad \quad \quad \quad \quad \quad \quad \text{ if } \epsilon \in \left [\frac{1}{2} - 3 \beta_\gamma ,1 \right )\\
        \sqrt{V(\Gamma) - |\mathcal{A}| \gamma \nu_{\max} }\, \Phi^{-1}\left(\epsilon + 3 \beta_\gamma  \right) + \frac{\log 2}{\sqrt{n}}  & \\
        \quad \quad \quad \quad \quad \quad \quad \quad \quad \quad \text{ if } \epsilon \in \left (0, \frac{1}{2} - 3 \beta_\gamma \right ). \label{myrval}
    \end{cases}  
\end{align}

Note that the upper bound in $(\ref{userhere})$ holds for any $r$. For a given error probability $\epsilon \in (0, 1)$, we choose $r$ according to $(\ref{myrval})$. Then using $(\ref{myrval})$ in $(\ref{userhere})$, we obtain for any given $\epsilon \in (0, 1)$ that  
\begin{align}
    P_\gamma\left( \sum_{i=1}^n Z_i > \sqrt{n} r - \log(2)   \right) &\leq 1 - \epsilon - 2\beta_\gamma. \label{firstterm} 
\end{align}
$(\ref{firstterm})$ provides an upper bound to the first term in $(\ref{q})$. 

We now upper bound the second term in $(\ref{q})$.

Using again the choice of $q$ in $(\ref{choiceq})$, we have 
\begin{align}
    &\sum_{t: T^n_{\mathcal{A}}(t) \subset \mathcal{P}_n^\gamma} P_t\left( \frac{W(Y^n|X^n)}{q(Y^n)} > \rho  \right) \notag \\
    &\leq \sum_{t: T^n_{\mathcal{A}}(t) \subset \mathcal{P}_n^\gamma} P_t\left( \frac{W(Y^n|X^n)}{ \prod_{i=1}^n q_t(y_i) } > \frac{\rho}{2 |\mathcal{P}_n(\mathcal{A})|}  \right) \notag \\
    &\leq \sum_{t: T^n_{\mathcal{A}}(t) \subset \mathcal{P}_n^\gamma} P_t\Bigg ( \sum_{i=1}^n \log \frac{W(Y_i|X_i)}{q_t(Y_i)}  > n C(\Gamma) + \sqrt{n} r - \mbox{} \notag \\
    & \quad \quad \quad \quad \quad \quad \quad \quad \quad \quad \quad \quad \quad \log 2(n+1)^{|\mathcal{A}|}   \Bigg ) \notag \\
    &\stackrel{(a)}{=} \sum_{t: T^n_{\mathcal{A}}(t) \subset \mathcal{P}_n^\gamma} W\Bigg ( \sum_{i=1}^n \log \frac{W(Y_i|x_{t,i})}{q_t(Y_i)}  > n C(\Gamma) + \sqrt{n} r - \mbox{} \notag \\
    & \quad \quad \quad \quad \quad \quad \quad \quad \quad \quad \quad \quad \quad\log 2(n+1)^{|\mathcal{A}|}   \Bigg), \label{typeetype}
\end{align}
where in equality $(a)$, $(x_{t,1}, \ldots, x_{t,n})$ is any arbitrary sequence from the type class $T^n_{\mathcal{A}}(t)$. Equality $(a)$ holds because under the probability measure $P_t$, $t(X^n) = t $ a.s. (see Remark \ref{remarktype}) and the distribution of 
$$\sum_{i=1}^n \log \frac{W(Y_i|X_i)}{q_t(Y_i)}$$
depends on $X^n$ only through its type. 
Continuing from $(\ref{typeetype})$, we have 

\begin{align}
    &\sum_{t: T^n_{\mathcal{A}}(t) \subset \mathcal{P}_n^\gamma} P_t\left( \frac{W(Y^n|X^n)}{q(Y^n)} > \rho  \right) \notag \\
    &\leq \sum_{ t: T^n_{\mathcal{A}}(t) \subset \mathcal{P}_n^\gamma} W \Bigg ( \sum_{i=1}^n \left [  \log \frac{W(Y_{i}|x_{t,i})}{q_t(Y_{i})} - \mbox{} \right .\notag \\
    & \left . \quad \quad \quad  \mathbb{E}_W \left [\log \frac{W(Y|x_{t,i})}{q_t(Y)}  \right ] \right ]  > n \left [ C(\Gamma) - I(t,W) \right]  + \mbox{} \notag \\
    & \quad \quad \quad \quad \quad \quad \quad \quad \sqrt{n} r - \log 2(n+1)^{|\mathcal{A}|}  \Bigg )\notag\\
    &\leq \sum_{t: T^n_{\mathcal{A}}(t) \subset \mathcal{P}_n^\gamma} W \left( \sum_{i=1}^n  \left [  \log \frac{W(Y_{i}|x_{t,i})}{q_t(Y_{i})} - \mbox{} \right. \right. \notag  \\
    & \left . \left . \quad \quad \quad \quad \quad \quad \quad \quad \mathbb{E}_W \left [\log \frac{W(Y|x_{t,i})}{q_t(Y)}  \right ] \right ]  > n\frac{K}{2} \right) \label{recycle}
\end{align}
where the last inequality holds for sufficiently large $n$ because $r$, as defined in $(\ref{myrval})$, is an $O(1)$ term, and from the construction of the set $\mathcal{P}_n^\gamma$, we have 
$$\inf_{t: T^n_{\mathcal{A}}(t) \subset \mathcal{P}_n^\gamma} \,d_W(t) \ge \gamma > 0$$
which implies
$$ \inf_{t: T^n_{\mathcal{A}}(t) \subset \mathcal{P}_n^\gamma}\left [ C(\Gamma) - I(t, W) \right ] > K $$
for some constant $K > 0$. 

Let $i_{\max, t} := \max_{a, b: q_t(b)W(b|a) > 0} \big | \log \frac{ W(b|a)}{q_t(b)} \big |$. We now show that $i_{\max, t} \leq 2 \log n$ for all $t$. Let $W_{\min}:= \min_{a, b: W(b|a) > 0} W(b|a)$ and $q_{\min, t} := \min_{b:q_t(b) > 0} q_t(b)$. Then 
\begin{align*}
    q_{\min, t} &= \min_{b:q_t(b) > 0} \sum_{a \in \mathcal{A}} t(a) W(b|a)\\
    &\geq \min_{a, b: W(b|a) > 0} W(b|a) \min_{a:t(a) > 0} t(a)\\
    &= \frac{W_{\min}}{n}.
\end{align*}
Thus,
\begin{align*}
     i_{\max, t} &=  \max_{a, b: q_t(b)W(b|a) > 0} \big | \log \frac{ W(b|a)}{q_t(b)} \big |\\
    &\leq \max_{a, b: q_t(b)W(b|a) > 0} \big | \log W(b|a) \big | + \max_{b:q_t(b) > 0} \big | \log q_t(b) \big |\\
    &\leq \log \frac{n}{W_{\min}^2}\\
    &\leq 2 \log n
\end{align*}
for all sufficiently large $n$. 
Hence, we can use Azuma's inequality \cite[(33), p. 61]{azumaguy} to upper bound $(\ref{recycle})$, giving us  
\begin{align}
    & \sum_{t: T^n_{\mathcal{A}}(t) \subset \mathcal{P}_n^\gamma} P_t\left( \frac{W(Y^n|X^n)}{q(Y^n)} > \rho  \right)  \notag\\
    &\leq (n + 1)^{|\mathcal{A}|} \exp \left( -\frac{n K^2}{128  \log^2 n } \right) \label{pn4}
\end{align}
which goes to zero as $n \to \infty$. 

Substituting the upper bounds $(\ref{firstterm})$ and $(\ref{pn4})$ in $(\ref{q})$, we obtain 
\begin{align*}
    &(F \circ W) \left( \frac{W(Y^n|X^n)}{q(Y^n)} > \exp\left( n C(\Gamma) + \sqrt{n} r \right) \right)\\
    &\leq 1 - \epsilon - \beta_\gamma  
\end{align*}
for sufficiently large $n$. Since the controller $F$ was arbitrary, we can apply Lemma \ref{qandp2} with $\rho = \exp\left( nC(\Gamma) + \sqrt{n}r \right)$, where $r$ is given in $(\ref{myrval})$, to obtain 
\begin{align}
    \log M^*_{\text{fb}}(n, \epsilon,  \Gamma) &\leq nC(\Gamma) + \sqrt{n} r - \log \beta_\gamma \notag  \\
    \frac{\log M^*_{\text{fb}}(n, \epsilon,  \Gamma) - n C(\Gamma) }{\sqrt{n}} &\leq r - \frac{\log \beta_\gamma}{\sqrt{n}} \notag \\
    \limsup_{n \to \infty} \frac{\log M^*_{\text{fb}}(n, \epsilon,  \Gamma) - n C(\Gamma) }{\sqrt{n}} &\leq r', \notag 
\end{align}
where $r'$ is obtained from the expression of $r$ in $(\ref{myrval})$ after taking the limit as $n \to \infty$, i.e., 
\begin{align*}
    &r' = \\
    &\begin{cases}
        \sqrt{V(\Gamma) + |\mathcal{A}| \gamma \nu_{\max} }\, \Phi^{-1}\left(\epsilon + 3 \beta_\gamma  \right)   & \text{ if } \epsilon \in \left [\frac{1}{2} - 3 \beta_\gamma ,1 \right )\\
        \sqrt{V(\Gamma) - |\mathcal{A}| \gamma \nu_{\max} }\, \Phi^{-1}\left(\epsilon + 3 \beta_\gamma  \right)   & \text{ if } \epsilon \in \left (0, \frac{1}{2} - 3 \beta_\gamma \right ). \label{myrval2}
    \end{cases}  
\end{align*}
Finally, since $\frac{V(\Gamma)}{4|\mathcal{A}| \nu_{\max}} > \gamma > 0$ was arbitrary, we can take $\gamma$ and $\beta_\gamma$ arbitrarily small, giving us the converse result 
\begin{align*}
    \limsup_{n \to \infty} \frac{\log M^*_{\text{fb}}(n, \epsilon,  \Gamma) - n C(\Gamma) }{\sqrt{n}} \leq \sqrt{V(\Gamma)} \Phi^{-1}(\epsilon). 
\end{align*}
Since this matches the optimal non-feedback SOCR of simple dispersion DMCs with the a.s. cost constraint, we have 
\begin{align}
    \lim_{n \to \infty} \frac{\log M^*_{\text{fb}}(n, \epsilon,  \Gamma) - n C(\Gamma) }{\sqrt{n}} = \sqrt{V(\Gamma)} \Phi^{-1}(\epsilon) 
\end{align}
for simple-dispersion DMCs with the a.s. cost constraint. 

\appendices

\section{Proof of Lemma \ref{restatedlemma} \label{restatedlemma_proof}}

We have 
\begin{align*}
    &\sum_{i=1}^n \mathbb{E}_\gamma\left [ Z_i^2 | \mathcal{G}_{i-1} \right] \\
    &= \sum_{i=1}^n \mathbb{E}_\gamma\left [ \left(i(X_i,Y_i)- \mathbb{E}\left [i(X_i, Y_i)|X_{i} \right] \right)^2 | \mathcal{G}_{i-1} \right]\\
    &= \sum_{i=1}^n \mathbb{E}_\gamma\left [ \left(i(X_i,Y_i)- \mathbb{E}\left [i(X_i, Y_i)|X_{i} \right] \right)^2 | X_{i} \right]\\
    &= \sum_{i=1}^n \text{Var}\left( i(X_i, Y_i) | X_i \right)\\
    &= \sum_{i=1}^n \sum_{a \in \mathcal{A}} \mathds{1}\left(X_i = a \right) \nu_a\\
    &= n \sum_{a \in \mathcal{A}}  P_{X^n}(a) \nu_a. 
\end{align*}
From Remark \ref{ha}, since $d_W(t(X^n)) \leq \gamma$ a.s., there exists a $\tilde{P} \in \Pi_{W, \Gamma}^*$ such that $||t(X^n) - \tilde{P}||_1 \leq 2 \gamma$. Hence,  
\begin{align*}
     n V(\Gamma) - 2n \gamma \nu_{\max}  &\leq \sum_{i=1}^n \mathbb{E}_\gamma\left [ Z_i^2 | \mathcal{G}_{i-1} \right] \leq n V(\Gamma) + 2n \gamma \nu_{\max}
\end{align*}
a.s., where we used the fact that $W$ is simple-dispersion at the cost $\Gamma$. Furthermore, since $\mathcal{F}_{i-1} \subset \mathcal{G}_{i-1}$,
\begin{align*}
    n V(\Gamma) - 2n \gamma \nu_{\max}  &\leq \sum_{i=1}^n \mathbb{E}_\gamma\left [ Z_i^2 | \mathcal{F}_{i-1} \right] \leq n V(\Gamma) +2 n \gamma \nu_{\max}
\end{align*}
a.s.
and 
\begin{align*}
    n V(\Gamma) - 2n \gamma \nu_{\max}  &\leq \sum_{i=1}^n \mathbb{E}_\gamma\left [ Z_i^2 \right] \leq n V(\Gamma) + 2n \gamma \nu_{\max}.
\end{align*}
Finally, we have 
\begin{align*}
    &\Bigg | \frac{\sum_{i=1}^n \mathbb{E}_\gamma[Z_i^2| \mathcal{F}_{i-1}]}{\sum_{i=1}^n \mathbb{E}_\gamma[Z
    _i^2]} - 1  \Bigg |\\
    &\leq \Bigg | \frac{ V(\Gamma) + 2 \gamma \nu_{\max}}{ V(\Gamma) -2  \gamma \nu_{\max}} - 1  \Bigg |\\
    &= \Bigg | \frac{ 4 \gamma \nu_{\max}}{ V(\Gamma) - 2 \gamma \nu_{\max}}   \Bigg |\\
    &\leq \frac{ 8 \gamma \nu_{\max}}{V(\Gamma)},
\end{align*}
assuming $\gamma \leq \frac{V(\Gamma)}{4 \nu_{\max}}$.

\section*{Acknowledgment}

This research was supported by the US National Science
Foundation under grant CCF-1956192.

\bibliographystyle{IEEEtran}
%




\end{document}